\documentclass[prb,twocolumn,superscriptaddress,showpacs,floatfix]{revtex4}
\usepackage{graphicx}
\begin{document}  
\title{Reliability of the Heitler-London approach for the exchange coupling\\
between electrons in semiconductor nanostructures}
\author{A. L. Saraiva}
\affiliation{Instituto de F\'{\i}sica, Universidade Federal do Rio de
Janeiro, Caixa Postal 68528, 21941-972 Rio de Janeiro, Brazil}
\author{M. J. Calder\'on}
\affiliation{Instituto de Ciencia de Materiales de Madrid (CSIC), Cantoblanco,
28049 Madrid, Spain}
\author{Belita Koiller}
\affiliation{Instituto de F\'{\i}sica, Universidade Federal do Rio de
Janeiro, Caixa Postal 68528, 21941-972 Rio de Janeiro, Brazil}
\date{\today}
\begin{abstract}
We calculate the exchange coupling $J$ between electrons in a
double-well potential in a two-dimensional semiconductor environment
within the Heitler-London (HL) approach. 
Two functional forms are considered for the single-well potential.
We show that by choosing an appropriate and relatively simple single-electron
variational wave function it is possible, \emph{within the HL approach}, to significantly improve the
estimates for $J$. In all cases the present scheme overcomes the artifacts and limitations 
at short interdot distances, previously attributed to the HL method, where unphysical
triplet ground states have been found, and leads to an overall agreement
with analytic interpolated  expressions for $J$ obtained for a donor-type model
potential.
\end{abstract}
\maketitle

For quantum computer architectures based on electron spins,~\cite{loss98,kane98,vrijen00,levy01,friesen03} two-qubit logical gates can be implemented by entangling electrons bound to neighboring potential wells through a transient exchange coupling.~\cite{loss98} 
Accurate evaluation of this coupling---namely, the difference in energy between triplet and singlet two-electron ground-state configurations---requires numerically intensive calculations which are highly dependent on the physical parameters of the system. Such parameters are in general not precisely known, and for this reason the Heitler-London (HL) method~\cite{heitler27, herring62} is of special interest in providing intuitive insight and computational simplicity 
for prospective estimations of the exchange coupling strength over a wide range of potential-related parameters. 
This is particularly important for gated quantum dots in which the exact form of the potentials, produced by electrodes over a two-dimensional electron gas (2DEG), has to be modeled. Also, within the HL approach, it is in principle possible to explore the exchange energy at large interwell distances, which would require great computational effort in more accurate methods. 

Prospective model calculations based on the HL method have led to an inaccurate exchange coupling asymptotic (large interdot distances) decay and to unphysical artifacts such as predicting a triplet ground state at short interwell distances (negative-$J$ anomaly).~\cite{burkard99,CKDexch}
It is usually argued that these problems are due to limitations in the HL approach,\cite{burkard99,CKDexch} and it has been suggested that they could only be overcome within more complex quantum chemistry or numerical
methods.~\cite{hu00,sousa01,hu01,scarola05,helle05,yannouleas02,anisimov00,stopa,pedersen07} 
We show here that the limitations of previous studies are mainly due to the choice of the single-particle wave functions rather than being intrinsic to the HL method.

Within the single-valley effective mass approximation, 
the Hamiltonian for an electron trapped in a single cylindrically symmetric well potential in a 2D semiconductor environment is given by $h
(\rho) = -\frac{\partial^2}{\partial \rho^2} + V(\rho)$, where $\rho=\sqrt{x^2+y^2}$ is the distance from the well minimum, and the energies
and distances are given in atomic units Ry$^*={{m_\perp e^4}/{2\hbar^2\varepsilon^2}}$ and $a^*={{\hbar^2\varepsilon}/{m_\perp e^2}}$,
respectively. Here the semiconductor is characterized by $m_\perp$, the transverse effective mass, and by the dielectric constant $\varepsilon$.
For Si, the experimental values of $m_\perp$ and $\varepsilon$ lead to $a^*_{Si}=3.157$ nm and Ry$^*_{Si}=19.98$ meV. For GaAs-based quantum
dots, these values are $a^*_{GaAs}=10.343$ nm and Ry$^*_{GaAs}=5.31$ meV. The first term in the Hamiltonian $h$ refers to the kinetic energy of the electron, and the second term is the model potential, which depends on the architecture of the physical system involved.

The confining potential produced by a simple shallow donor (e.g., P) positioned a distance $d$ from an interface with a barrier is given by
\begin{equation}
V(\rho) = -\frac{2}{\sqrt{\rho^2+d^2}}~.
\label{eq:Vdonor}
\end{equation}
This does not include the effect of the image charge in the Si/SiO$_2$ architecture, which could be accounted for straightforwardly by multiplying the potential by 
$2  \varepsilon_{Si}/(\varepsilon_{Si}+\varepsilon_{SiO_2}) \approx 1.5$. Near the potential minimum we get $V(\rho\approx 0) = -2/d+\rho^2/d^3$, defining the parabolic approximation for this model potential.~\cite{burkard99}

For gated quantum dots, several model potentials have been adopted in the literature. In particular, the functional form in 
Eq.~(\ref{eq:Vdonor}) yields a nearly harmonic confinement around the dot center and asymptotically follows a Coulombic decay at large distances from the center [$V(\rho\rightarrow\infty)\approx -2/\rho$].
Previous studies of gated quantum dots (e.g., Refs.~\onlinecite{hu00,hu01,sousa01} and~\onlinecite{szafran04}) propose a faster decay for the quantum dot confining potential, given by a Gaussian form
\begin{equation}
V^\prime(\rho) = -V_0\,e^{-\rho^2/L^2}~.
\label{eq:Vqd}
\end{equation}
We also consider this model potential below.

\begin{figure}
\begin{center}
\resizebox{85mm}{!}{\includegraphics{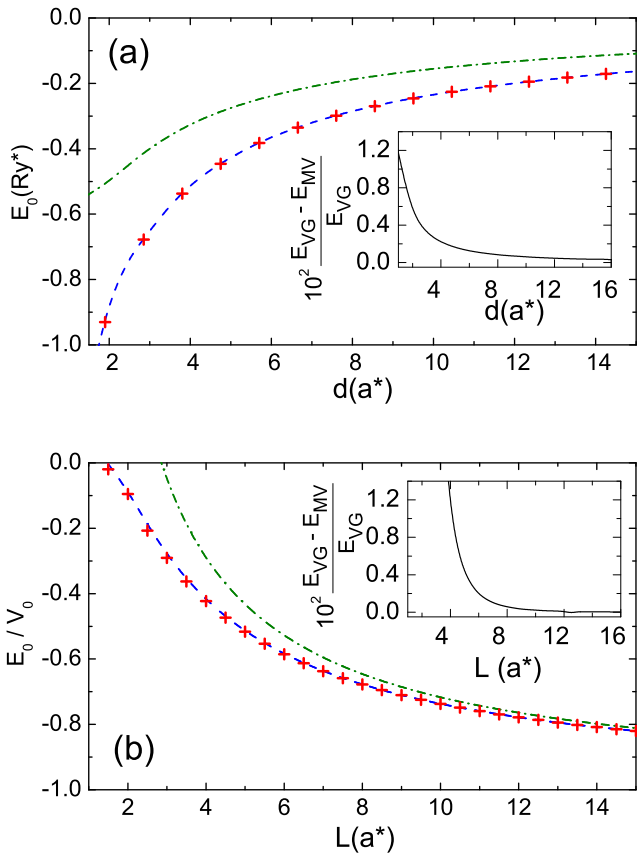}}
\caption{\label{fig:energy}(Color online) Ground-state energies obtained for the three approximations considered here. 
As the dot-dashed lines, the much larger energies obtained through the parabolic Gaussian (PG) scheme. As the dashed lines, the energies for the
variational Gaussian (VG), which are very close to those calculated through the more general matched variational (MV), given by the crosses.
The insets show the percent reduction on the energy from VG to MV. Results in (a) correspond to the potential $V(\rho)$ in Eq.~(\ref{eq:Vdonor}) while results for $V^\prime(\rho)$ in Eq.~(\ref{eq:Vqd}) are given in (b).}
\end{center}
\end{figure}

For $d=0$ the Hamiltonian with $V(\rho)$ in Eq.~(\ref{eq:Vdonor}) reduces to that of a 2D hydrogen atom,\cite{ponomarev992,CKDexch} so the exact ground-state solution is  $\phi_0 = \alpha_{0}\exp(-\alpha_{0}\rho)/\sqrt{2\pi}$,  with $\alpha_0^{-1}=0.5a^*$, and the energy is $E(d=0)=-4$Ry*. In the general case ($d \ne 0$), some approximation must be adopted. The simplest approach consists of using the exact ground-state wave function of the parabolic potentials defined in the paragraph following Eq.~(\ref{eq:Vdonor}). We call this approach the parabolic Gaussian (PG), and the ground state has the functional form $\phi_{PG} = \beta_{PG}/\sqrt{\pi} \, \exp\left({-\beta_{PG}^2 \rho^2/2} \right)$, where $\beta_{PG}=(2/d^3)^{1/4}$, and the energies are given by $E_{PG}=2  \beta^2-2/d$. 
A better approach\cite{calderonPRL06} is to treat $\beta$ as a variational parameter [variational Gaussian(VG) wave function approach] keeping the exact form of the potential. This leads to significant energy reduction as compared to PG,\cite{calderonPRL06} 
as shown in Fig.~\ref{fig:energy}(a).

\begin{figure}
\begin{center}
\resizebox{85mm}{!}{\includegraphics{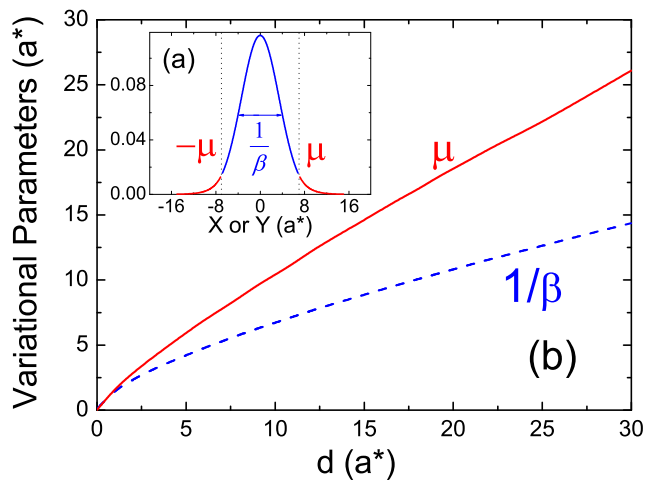}}
\caption{\label{fig:param} (Color online) (a) Illustration of $|\phi_{MV}|^2$ for $d=6a$* [see Eq.~(\ref{eq:MV})]. (b) The optimized variational parameters for MV, $\mu$ and $1/\beta$, are shown here to be of the same order of magnitude.
The given results correspond to a single well of the form $V(\rho)$ as in Eq.~(\ref{eq:Vdonor}).}
\end{center}
\end{figure}

The VG form of the wave function is well suited close to the potential minimum (where the parabolic fit is more appropriate). At large distances from the well minimum, however, the potential
is finite and we expect the ground-state wave function to decay exponentially rather than as a Gaussian. The exponential decay is expected for any electronic bound state ($E_0$) in a potential that goes to zero at large distances, contrary to the Gaussian decay, characteristic of the infinitely confining parabolic well. This argument also applies to the Gaussian model potential $V^\prime(\rho)$ in Eq.~(\ref{eq:Vqd}). 
To correctly take into account the form of the potential at short $\rho\rightarrow 0$ and long $\rho\rightarrow \infty$ distances, we define a variational wave function by parts: (i) a Gaussian behavior up to some distance $\rho < \mu$, suitable for a nearly parabolic attractive potential, and (ii) an exponential decay for $\rho>\mu$, adequate for a finite potential. This matched variational (MV)  wave function [illustrated in Fig.~\ref{fig:param}(a)] is written
\begin{equation}
\phi_{MV}(\rho) = \left\{ \begin{array}{ll}
 A_1 \exp{\left(-\frac{\beta^2 \rho^2}{2}\right)} &\mbox{ if $\rho<\mu$},\\
 A_2 \exp{\left(-\frac{\alpha \rho}{2}\right)} &\mbox{ if $\rho > \mu$},
       \end{array} \right.
\label{eq:MV}
\end{equation}
and involves the parameters $A_1$, $A_2$, $\beta$, $\alpha$, and $\mu$. From the continuity of the wave function and of its first derivative at $\rho=\mu$, and normalization, two independent variational parameters are left, which we choose to be $\beta$ and $\mu$. These are obtained (Fig.~\ref{fig:param}) by minimizing the energy expectation value. 
The parameter $\alpha$ is obtained through the constraint $\alpha=\beta^2\mu$.
Note that the MV wave function reproduces VG for $\mu\gg1/\beta$. 

The variational principle regards only the value
of the energy, drawing no conclusions on the validity of the trial
wave function.~\cite{cohen} Since the main contribution to the expectation value of the energy is given by the $\rho$ values closer to the potential minimum, the more general MV does not lower the energy much (typically less than $1\%$) with respect to the VG, as shown in the insets in Fig.~\ref{fig:energy}. 

In general, the optimized variational parameters $\mu$ and $1/\beta$ are of the same order
of magnitude, as shown, for example, in Fig.~\ref{fig:param}(b). This means that, although the
energy gain of MV with respect to the VG is not very significative, the form of the wave
functions is different, particularly at the tails.
The functional form of the wave function decay is relevant for the exchange coupling, as $J$ depends very
strongly on the neighboring wave-function overlap. Around the
potential minimum, however, the difference between the VG and MV wave functions is not
so important, since $\beta_{VG}\approx\beta_{MV}$ for $d\gtrsim 3 a^*$.

We write the Hamiltonian for a two-electron system in a double well as 
\begin{equation}
H(1,2) = \sum_{i=1,2}\left[-\nabla_i^2 +V_{DW}(i)\right]+\frac{2}{r_{12}},
\label{eq:HDW2}
\end{equation}
where $i$ labels each electron and the last term is the electron-electron Coulomb repulsion. 
The double-well potential $V_{DW}(i)$ consists of identical wells a distance $R$ from each other,
\begin{equation}
V_{DW}(i) = V\left(x_i-\frac{R}{2}, y_i\right) + V \left(x_i+\frac{R}{2}, y_i\right).
\label{eq:VDW}
\end{equation}
We denote as $\left|{\phi_A(i)}\right\rangle$ and $\left|{\phi_B(i)}\right\rangle$ the one-electron orbitals centered at $+R/2$ and $-R/2$, respectively.
Within the HL approach, the exchange coupling is calculated as $J=E_{triplet}-E_{singlet}$ with the singlet and triplet wave functions written as the symmetric and antisymmetric combinations of the one-electron orbitals.\cite{calderon07}

Anomalies in the HL approach within the PG and VG have been pointed out in the literature. In the presence of a finite interwell barrier, $J$ should decay exponentially~\cite{herring64} for $R\rightarrow \infty$.
This cannot be obtained through PG or VG orbitals, as both have a Gaussian decay at all distances. 
Besides the failure in reproducing the expected asymptotic behavior, the PG and VG also lead to unphysical~\cite{lieb62} negative values of the coupling at small $R$.~\cite{burkard99,CKDexch} 

\begin{figure}
\begin{center}
\resizebox{85mm}{!}{\includegraphics{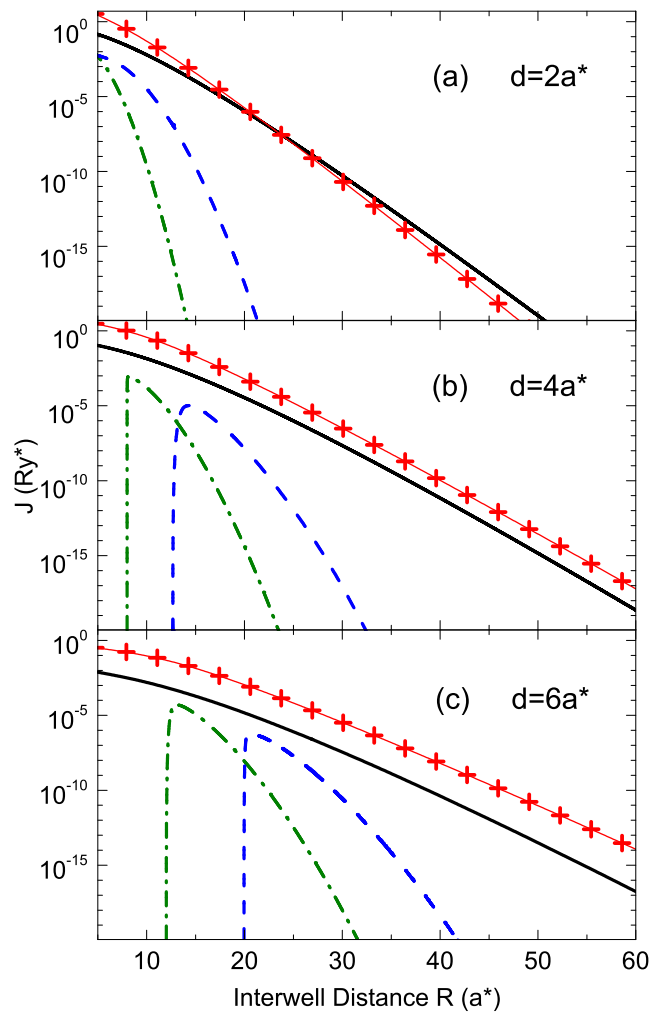}}
\caption{\label{fig:exchange}(Color online) Exchange coupling calculated for the indicated values of the geometric parameter $d$ for the quantum well potential $V(\rho)$ in Eq.~(\ref{eq:Vdonor}). The dashed lines and crosses are obtained within the HL approach: Dot-dashed lines for the PG approximation, dashed lines for VG, and crosses for MV. The solid line corresponds to the expression obtained in Ref.~\onlinecite{ponomarev992} from the asymptotic behavior for $R\rightarrow 0$ and $R\rightarrow \infty$. The negative-$J$ anomaly is indicated by the unphysical sharp drop in the PG and VG curves at small $R$ values. At large $R$, deviations of PG and VG from the correct asymptotic decay increase exponentially.}
\end{center}
\end{figure}

In Fig.~\ref{fig:exchange}, we show the calculated exchange coupling for the three forms considered for $\phi$. We also show the trend of the dependence of $J$ upon the geometrical parameter $d$ by showing the results for $d = 2a$*, $4a$*, and $6a$*. There are two main results to stress here: (i) while the PG and VG approaches give unphysical negative values for the exchange at short distances $R$, the MV wave function gives a positive $J$ for the entire range of parameters studied;\cite{foot3}
(ii) at large $R$, MV gives the expected exponential decay for $J(R)$ (as inferred from the linear behavior in the log scale of the figure). This decay is much slower than the one for VG and PG, leading to differences of many orders of magnitude in $J$ among the different approaches.

The ``out-of-plane donor'' type of potential we are considering was previously discussed in
Ref.~\onlinecite{ponomarev992}, where analytic expressions for the exchange integral were interpolated from asymptotic methods which are precise for $R\rightarrow0$ and $R\rightarrow\infty$. This result is compared to ours in Fig.~\ref{fig:exchange}, where we show that MV is in overall agreement with the interpolated expression. 

\begin{figure}
\begin{center}
\resizebox{85mm}{!}{\includegraphics{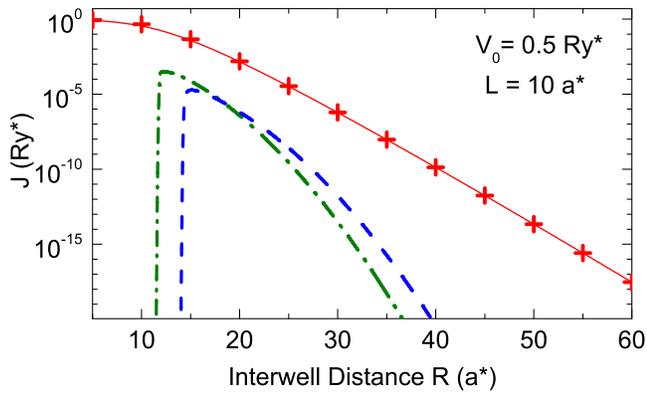}}
\caption{\label{fig:Jpotgauss}(Color online) Exchange energies calculated for the potential $V^\prime(\rho)$ in Eq.~(\ref{eq:Vqd}), with $V_0$ = 0.5Ry$^*$ and $L$ = 10$a^*$. The general behavior and trends here are equivalent to those in Fig.~\ref{fig:exchange}.
}
\end{center}
\end{figure}

We expect these results to remain valid for general additive double-well models, as long as there is a \emph{finite} barrier between the two wells. Indeed, for the Gaussian potential well given in Eq.~(\ref{eq:Vqd}), we find that MV still minimizes the energy expectation value [Fig.~\ref{fig:energy}(b)]. The calculated exchange coupling, illustrated in Fig.~\ref{fig:Jpotgauss}, shows the same trends as obtained for the donor-type potential, Eq.~(\ref{eq:Vdonor}), suggesting the general validity of our arguments for different types of potential wells.

The HL approach is a simple tool to estimate exchange coupling between electron spins in artificial and real molecules. However, physical inaccuracies reported in the past---namely, a triplet ground state predicted for small interdot distances and Gaussian instead of exponential asymptotic decay at large values of $R$ (Ref.~\onlinecite{CKDexch})---severely restrict its applicability. 
For applications in qubits, the small-$R$ behavior of $J$ would be the most relevant, since strong enough coupling is needed to guarantee fast gating times. On the other hand, reliable estimates for the large-$R$ coupling must also be available, as residual interactions between nominally noninteracting qubits constitute an additional source of errors affecting the desired quantum computer operations. 
Here we showed that these shortcomings are not intrinsic to the HL approach and may be attributed to the usual Gaussian form adopted for the one-electron single-well orbital. Previous studies\cite{yannouleas02, pedersen07} establish that, within a Gaussian basis set for the single-particle orbitals, overcoming the mentioned limitations requires at least 70 molecular orbitals to be considered in solving the two-particle problem.  
We have shown that a variational single-electron orbital with appropriate
long- and short-distance functional forms within the HL approach leads to a singlet ground state at short $R$ and to the correct exponential decay at large $R$. No quantitative accuracy is attempted here, as only the lowest-energy single-electron orbital is considered, and even so it is not the exact ground state for the single-dot potential. 

Some of us have recently proposed to perform exchange operations at the Si/SiO$_2$ interface between electrons laterally bound to donors a distance $d$ from the interface.~\cite{calderon07,calderon062} The form of the potential assumed here applies exactly to this problem (except for a 
multiplicative factor related to the image charge, as mentioned previously). The exchange coupling was calculated within the HL method through the VG approach for the wave function,~\cite{calderon07} leading to small values of the exchange, $\lesssim 10^{-3}$ meV $\sim 10^{-5} $Ry$^*$. The results reported here in Fig.~\ref{fig:exchange} imply that the actual values of the exchange that can be achieved at the Si/SiO$_2$ interface are orders of magnitude larger and therefore would require much shorter gating times than previously expected.

This work was partially supported in Brazil by CNPq, FAPERJ, FUJB, CAPES, and Millenium Institute of Nanotechnology--MCT. M.J.C. acknowledges support from Programa Ram\'on y Cajal (MEC, Spain).
\bibliography{hlparts}

\begin{thebibliography}{27}
\bibitem[{foo()}]{foot3}
\bibinfo{note}{Our scheme is not capable or intended to overcome the
  negative-$J$ HL anomaly due to the singular nature of the Coulomb potential
  (discussed in Ref.~\onlinecite{herring62}), leading to a logarithmic
  dependence of the electron-electron repulsion energy with $R$, which becomes
  the dominant term at very large $R$.}
\end{thebibliography}

\end{document}